\documentclass[twocolumn]{aastex6}
 
\usepackage{color}
\usepackage{epsfig}

\slugcomment{submitted to the Astrophysical Journal}
 
\shorttitle{ASPECS: spectral line intensity mapping at millimeter
wavelengths and CMB spectral distortions}

\shortauthors{Carilli et al.}
 
\begin{document}

\title{The ALMA Spectroscopic Survey in the Hubble Ultra Deep Field:
implications for spectral line intensity mapping at millimeter
wavelengths and CMB spectral distortions}

\author{C.L. Carilli\altaffilmark{1,2}, J. Chluba\altaffilmark{3},
R. Decarli\altaffilmark{4}, F. Walter\altaffilmark{4,1},
M. Aravena\altaffilmark{5}, J. Wagg\altaffilmark{6},
G. Popping\altaffilmark{7}, P. Cortes\altaffilmark{8,9},
J. Hodge\altaffilmark{10}, A. Weiss\altaffilmark{11},
F. Bertoldi\altaffilmark{12}, D. Riechers\altaffilmark{13}}

\altaffiltext{1}{National Radio Astronomy Observatory, P. O. Box 0,
Socorro, NM 87801, USA; ccarilli@aoc.nrao.edu} 
\altaffiltext{2}{Astrophysics Group, Cavendish Laboratory, JJ Thomson Avenue,
Cambridge CB3 0HE, UK}
\altaffiltext{3}{Jodrell Bank Centre for Astrophysics, University of Manchester, 
Oxford Road, M13 9PL, UK}
\altaffiltext{4}{Max-Planck Institute for Astronomy, D-69117 Heidelberg, Germany}
\altaffiltext{5}{Nucleo de Astronomia, Facultad de Ingenieria, Universidad 
Diego Portales, Av. Ejercito 441, Santiago, Chile}
\altaffiltext{6}{Square Kilometre Array Organisation, Lower Withington, Cheshire, UK}
\altaffiltext{7}{European Southern Observatory, Karl-Schwarzschild-Strasse 2, 
85748, Garching, Germany}
\altaffiltext{8}{Joint ALMA Observatory - ESO, Av. Alonso de Cordova,
3104, Santiago, Chile}
\altaffiltext{9}{National Radio Astronomy Observatory, 520 Edgemont
Rd, Charlottesville, VA, 22903, USA}
\altaffiltext{10}{Leiden Observatory, Leiden University, Niels Bohrweg 2,
NL2333 RA Leiden, The Netherlands}
\altaffiltext{11}{Max-Planck-Institut f\"ur Radioastronomie, Auf dem H\"ugel
69, 53121 Bonn, Germany}
\altaffiltext{12}{Argelander Institute for Astronomy, University of Bonn,
Auf dem H\"ugel 71, 53121 Bonn, Germany}
\altaffiltext{13}{Cornell University, 220 Space Sciences Building, Ithaca,
NY 14853, USA}

\altaffiltext{$\star$}{ALMA is a partnership of ESO (representing its
member states), NSF (USA) and NINS (Japan), together with NRC (Canada)
and NSC and ASIAA (Taiwan), in cooperation with the Republic of Chile. 
The National Radio Astronomy Observatory is a facility of 
the National Science Foundation operated under cooperative agreement by 
Associated Universities, Inc.
}

\begin{abstract}

We present direct estimates of the mean sky brightness temperature in
observing bands around 99GHz and 242GHz due to line emission from
distant galaxies.  These values are calculated from the summed line
emission observed in a blind, deep survey for spectral line emission
from high redshift galaxies using ALMA (the 'ASPECS' survey).  In the
99 GHz band, the mean brightness will be dominated by rotational
transitions of CO from intermediate and high redshift galaxies. In the
242GHz band, the emission could be a combination of higher order CO
lines, and possibly [CII] 158$\mu$m line emission from very high
redshift galaxies ($z \sim 6$ to 7).  The mean line surface brightness
is a quantity that is relevant to measurements of spectral distortions
of the cosmic microwave background, and as a potential tool for
studying large-scale structures in the early Universe using intensity
mapping.  While the cosmic volume and the number of detections are
admittedly small, this pilot survey provides a direct measure of the
mean line surface brightness, independent of conversion factors,
excitation, or other galaxy formation model assumptions. The mean
surface brightness in the 99GHZ band is: $T_B = 0.94\pm 0.09\mu$K. In the
242GHz band, the mean brightness is: $T_B = 0.55\pm 0.033\mu$K.  These
should be interpreted as lower limits on the average sky signal, since
we only include lines detected individually in the blind survey, while
in a low resolution intensity mapping experiment, there will also be
the summed contribution from lower luminosity galaxies that cannot be
detected individually in the current blind survey.

\end{abstract}

\keywords{galaxies: formation, radio/FIR lines; submm: starbursts;
  physics: microwave background spectrum}

\section{Introduction}

Intensity mapping of the cumulative CO and other millimeter and
submillimeter line emission from early galaxies has been proposed as a
new means to probe very large-scale structures in the distant Universe
(Carilli 2011; Gong et al. 2011; Gong et al. 2012; Yue et al. 2015).
Intensity mapping entails low spatial and spectral resolution imaging
of the sky to obtain the mean brightness due to the cumulative
emission from myriad discrete cosmic sources. While interferometric
arrays like ALMA, the JVLA, and NOEMA, can detect CO and [CII]
158$\mu$m (and in cases of high luminosity sources, other lines), from
individual galaxies at high redshift, the fields of view are very
small, and the integration times are long. These telescopes are
inadequate for measuring the galaxy distribution on the very large
scales relevant to cosmological questions, such as the Baryon Acoustic
Oscillations at intermediate redshifts, or the large-scale
distribution of galaxies that reionize the Universe. The latter is of
particular interest for cross correlation studies with very wide
field, low-resolution HI 21cm images of the intergalactic medium
during cosmic reionization (Lidz et al. 2011).

The integrated millimeter and submillimieter line emission from early
galaxies has also been recognized as a possible significant
contaminant of measurements of the spectral and spatial fluctuations
of the cosmic microwave background (CMB; Righi et al. 2008a,b; Chluba
\& Sunyaev, 2012; de Zotti et al. 2015; Mashian et al. 2016). For
example, modeling suggests (Mashian et al. 2016) that the integrated
CO line emission could be significantly higher than the primordial
spectral distortions due to other cosmological effects (e.g., Chluba
\& Sunyaev, 2012; Sunyaev \& Khatri, 2013; Tashiro 2014), and may be
measurable with next generation instruments like the Primordial
Inflation Explorer (PIXIE; Kogut et al. 2014).\footnote{PIXIE is a
space observatory concept to map the CMB over the frequency range
30GHz to 6THz, one goal of which is to constrain the average CMB
energy spectrum with much greater accuracy than FIRAS.}

Numerous calculations have been done to predict the mean sky brightness due
to emission lines from CO at intermediate and high redshift, and [CII]
158$\mu$m emission at very high redshift (see section 2).  These
predictions are based on either empirical estimates using proxies for
the line emission, such as the cosmic star formation rate density, or
large scale cosmological simulations of galaxy formation, with recipes
to relate proxy measurements or simulated properties to line
luminosities.

In this paper, we present direct measurements of the summed line
luminosity from individual sources in bands around 99 GHz and 242 GHz.
These measurements are based on the ASPECS program, corresponding to a
broad band spectral line deep field of the UDF at 1.25mm and 3mm
(Decarli et al. 2016; Aravena et al. 2016a; Walter et al. 2016).  From
these measurement, we derive the mean brightness temperature at a 
given observing frequency due to high redshift galaxies. As a pilot
study with ALMA, the fields are necessarily small, and the number of
galaxies few. However, the measured quantity is direct: line emission
from early galaxies. Hence, no modeling or conversion factors are
required.

\section{Model predictions for the line brightness}

The dominant contribution to the integrated line brightness from high
redshift galaxies in the 99GHz band is due to rotational transitions
of CO from galaxies at intermediate to high redshift. Other molecular
tracers, such as the high dipole moment molecules like HCN and HCO$^+$,
are typically 10 times, or more, fainter than CO, while the atomic
fine structure lines would be from galaxies at improbable redshifts
($z \sim 20$; see Carilli \& Walter 2013). At 242GHz, the integrated
line brightness will be some combination of higher order CO lines from
intermediate and high redshift galaxies, plus a possible contribution
from [CII] 158$\mu$m line emission from galaxies at $z \sim 6$ to 7,
and other fine structure lines at lower redshift. We consider each in
turn.

Considering CO in the 99GHz band, predictions of the mean CO sky
brightness from early galaxies have taken two approaches. First is an
empirical use of the measured evolution of the cosmic star formation
rate density, and/or the cosmic FIR background, converted to CO
luminosity by adopting a CO-to-FIR or star formation rate conversion
factor (Lidz et al. 2011; Righi et al. 2008; de Zotti et al. 2015). A
related calculation is to consider the star formation rate density
required to reionize the neutral intergalactic medium at high
redshift, subsequently converted to CO luminosity using said
conversion factors (Carilli et al. 2011; Gong et al. 2011).  While
based on celestial measurements, these methods involve significant
uncertainties inherent in both the determination of the cosmic star
formation rate density, and more importantly, the assumed 'star
formation law' relating CO luminosity to FIR luminosity, or to star
formation rate (Kennicutt \& Evans 2012; Carilli \& Walter 2013).  The
latter may entail a dual conversion of star formation rate to total
gas mass, then total gas mass to CO luminosity. There is the
additional uncertainty in the assumed gas excitation when modeling the
contribution to the mean brightess at a given observing frequency from
different CO transitions from galaxies at different redshifts.

The second method for predicting the mean CO sky brightness is through
cosmological numerical simulations (Mashian et al. 2016; Gong et
al. 2011; Li et al. 2016). Such simulations can be normalized to
eg. an observed galaxy luminosity function at a given redshift,
although ultimately, even the most detailed simulations rely on
recipes to convert from simulated to observable quantities.  This is
particularly difficult in the case of tracer molecules, such as CO,
while also including their excitation state.

In summary, the predictions at around 100GHz for the mean brightness
from CO lines from intermediate and high redshift galaxies range from
1.5$\mu$K (Righi et al. 2008) to about 10$\mu$K (Marshian \& Loeb
2016).

Two recent observations have set upper limits to the CO brightness
from distant galaxies using CO intensity mapping. The first entailed a
cross correlation of WMAP images with maps of very large-scale
structures from the SDSS, namely the photometric quasar sample and the
luminous red galaxy sample (Pullen et al. 2013). The cross correlation
technique removes numerous systematic errors.  Pullen et al. estimate
upper limits to the mean brightness temperature of CO 1-0 or 2-1 of
about $10\mu$K in the 30 GHz to 90 GHz range. The second was an
interferometric measurement of the brightness fluctuations at 30GHz
using the Sunyaev-Zel'dovich Array (Keating et al. 2015). Keating et
al. quote an upper limit to the CO 1-0 mean brightness of $\sim 5\mu$K
from $z \sim 3$ galaxies.  Since both these measurements rely on
modeling of the spatial structure in the signal, they depend on
the assumed under-lying structural parameters.

Considering the 242GHz band, predictions also vary considerably. The
most detailed modeling to date, including analysis of lines from CO,
[CII], and [CI], is presented in Yue et al. (2015). They use large
scale cosmological simulations, plus physically motivated conversion
factors (Vallini et al. 2015; Pallottini et al. 2015), to derive the
line luminosities from early galaxies.  They predict a [CII] 158$\mu$m
brightness of $\sim 0.05\mu$K around 242GHz from $z \sim 6.5$
galaxies. At this frequency, they obtain a similar contribution from
the [CI] lines at rest frame frequencies of 492GHz and 809GHz, from
lower redshift galaxies. The dominant line contribution to the mean
brightness at 242GHz in their models comes from CO emission from
galaxies at intermediate to high redshift, for which they derive a
mean brightness of $\sim 0.45\mu$K.  Note that, in their models, the
[CII] contribution increases rapidly with increasing observing
frequency, to $\sim 0.4\mu$K at 316GHz (comparable to CO), due to
galaxies at $z \sim 5$.  Conversely, Gong et al. (2012) perform an
analytic calculation of the expected [CII] surface brightness based on
ISM physics and halo statistics, and predict a substantially larger
contribution of [CII] at 242 GHz of $\sim 0.3\mu$K.

\section{ASPECS: a blind search for millimeter line emission from high 
redshift galaxies}

The ALMA spectral deep field observations (ASPECS) and results are
described in Walter et al. (2016); Aravena et al. (2016a); Decarli et
al. (2016). In brief, we surveyed a $\sim 1$ arcmin$^2$ field in the
UDF to a 3-$\sigma$ depth of $\sim 0.05$ Jy km s$^{-1}$ (assuming line
widths of 200 km s$^{-1}$) over the frequency range 84--115 GHz (3
mm), and of $\sim 0.13$ Jy km s$^{-1}$ over the frequency range
212--272 GHz. In the analysis below, we adopt the mean frequencies
for each band, which are 99GHz and 242GHz.

The observations are sensitive to galaxies over a wide range in
redshift, depending on CO transition.  The typical CO luminosity
limits are $\sim 2\times10^9$ K km s$^{-1}$ pc$^2$ at 3mm and $6\times
10^8$ K km s$^{-1}$ pc$^2$ at 1mm at $1<z<3$.  The implied gas mass
limits will depend on which CO transition is being considered, at
which redshift, and depend critically on assumed CO excitation, in
particular for the high order transitions, since the total gas mass is
derived by extrapolation to low order. For reference, at the typical
redshifts and transitions of the detected galaxies, these limits imply
galaxies with gas masses of 3 to 10$\times 10^{9}$ M$_\odot$, for
a Galactic conversion factor of CO 1-0 luminosity to total molecular
gas mass. We discuss this point further in Section 4.1.
For [CII] at very high redshift ($z \sim 6.5$), our observations are
sensitive to galaxies with star formation rates $\ge 10$ M$_\odot$ year$^{-1}$,
using the de Looze et al. (2011) conversion. 

Line emitting galaxies were identified using multiple three
dimensional search algorithms, and a series of tests were made for
completeness and fidelity (Walter et al.2016).  Once a line
candidate was identified, a search was made for an optical or near-IR
counterpart. Decarli et al. (2016a) discuss how a given line is
identified as a specific CO transition. In some cases, the detection
(or the lack of detection) of multiple CO transitions over the broad
frequency range constrains the redshift determination. If an
optical/NIR counterpart is present, literature information on the
redshift of the source (via spectroscopy or SED fitting of the
photometry) was also used. The lack of an optical counterpart is used
to rule out low redshift interpretations in some cases.  Ultimately,
for some sources there can be ambiguity as to the transition in
question, and therefore the redshift.  This is dealt with via a
bootstrapping approach (see Decarli et al. 2016a).  However, in the
context of the analysis below, this is not an issue, since we 
simply sum all the lines detected in the blind survey in a give
observing band, independent of what transition and redshift the line
happens to be.

For completeness, Figure 1 shows a compilation of all the candidate
line detections in the survey, as presented in Walter et
al. (2016). There are a total of 21 candidate lines above
5$\sigma$. In 6 cases, line identifications are unequivocally
confirmed, through detection of other CO transitions, and/or an
optical galaxy with a spectroscopic redshift. For the rest of the
lines, extensive quantitative tests are made, and we only include
lines with a $> 60\%$ 'fidelity' rating. See Decarli et al. (2016a)
for more details on the statistical analysis, and Walter et al. (2016)
for total intensity CO images, and optical images, of all the
candidates. At this fidelilty level, we
expect to have roughly as many spurious detections as sources missing
from the survey due to noise fluctuations (Decarli et al. 2016a).

\begin{figure}
\centering 
\includegraphics[width=\columnwidth]{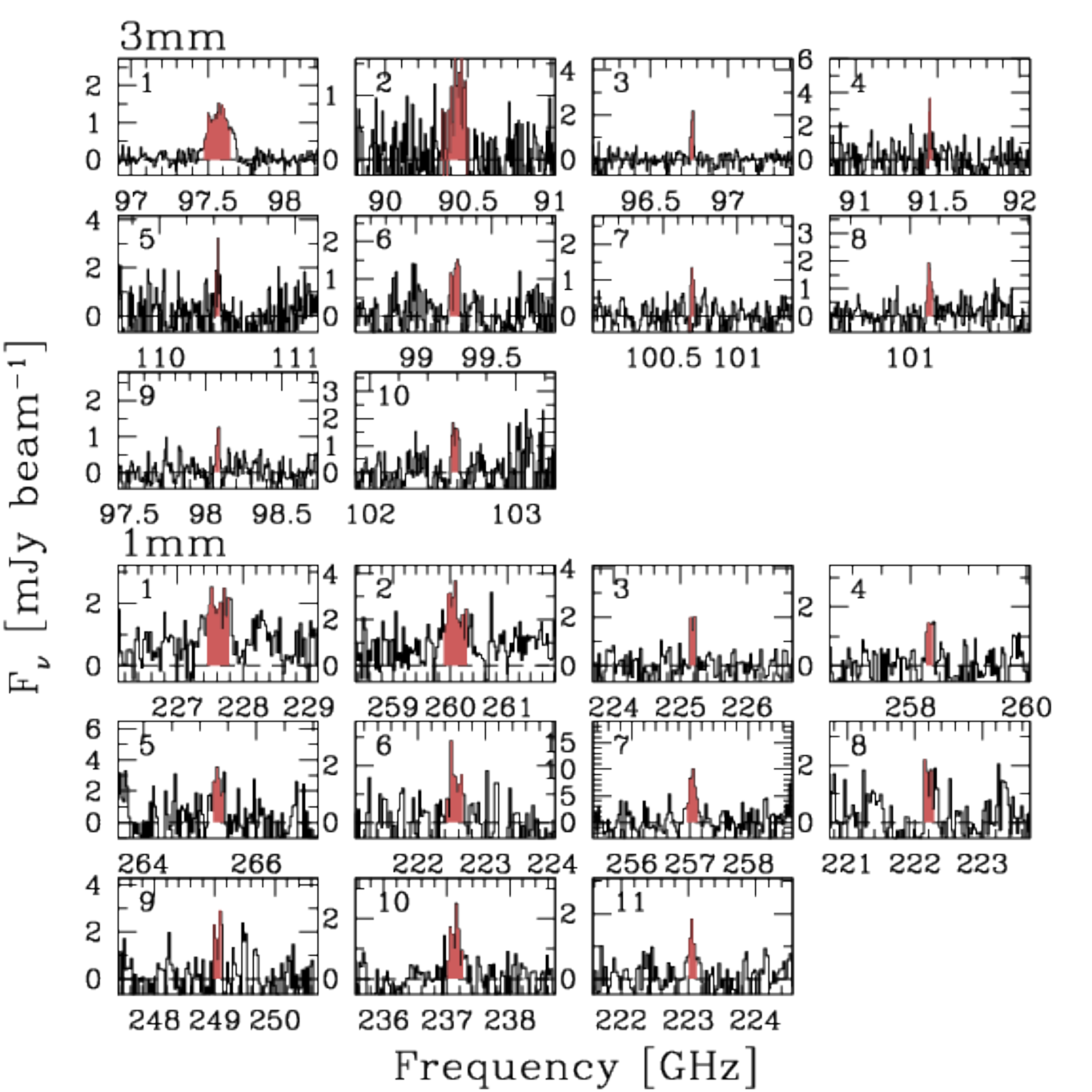}
\caption{Montage of all the $> 5\sigma$ line candidates from ASPECS 
blind survey (Walter et al.
2016). The red regions show the spectral region over which the signal-to-noise
was calculated (see Decarli et al. 2016).}
\label{fig:one}
\end{figure}

\section{Mean Brightness Temperatures}

The mean line brightness temperatures are calculated using an
empirical approach of summing the lines detected in the blind
survey. The process is simple: Table 2 in Walter et al. (2016) lists
the lines detected according to the blind search criteria outlined in
Section 3.  We sum the total flux from all the lines detected in a
given band, in Jy km s$^{-1}$, which is equivalent dimensionally to Jy
Hz or ergs s$^{-1}$ cm$^{-2}$. We then divide by the total band width
covered in the blind survey, which results in a mean flux density over
the full band and over the full field, $S_\nu$, in Jy. The mean
brightness temperature at the observed wavelength, $\lambda_{obs}$, is
then derived using the angular area of the field employed in the blind
search, $\Omega_f$, under the Rayleigh-Jeans appoximation: $T_B \sim
1360 ~ S_\nu ~ \lambda_{obs}^2 ~ \Omega_f^{-1} ~~ \rm K$, ~where
$\lambda_{obs}$ is in centimeters and $\Omega_f$ is in arcseconds$^2$.

For the 99GHz band, the total flux for the 10 lines detected in the
band is $2.53\pm 0.25$ Jy km s$^{-1} = (8.3\pm 0.08) \times 10^{5}$
Jy Hz.  The total bandwidth is 31GHz, so the mean flux density across
the band is: $S_\nu = (2.7 \pm 0.27) \times 10^{-5}$ Jy. The field
covered by the survey was 3600 arcsecond$^2$. Hence, the mean 
brightness temperature, $T_B = 0.94\pm 0.09\mu$K.

For the 242GHz band, the total flux for the 11 lines detected in the
band is $6.93\pm 0.42$ Jy km s$^{-1} = (5.6 \pm 0.34) \times 10^6$
Jy Hz.  The total bandwidth is 60GHz, so the mean flux density across
the band is: $S_\nu = (9.3 \pm 0.6) \times 10^{-5}$ Jy. The field
covered by the survey was also 3600 arcsecond$^2$. Hence, the mean
brightness temperature, $T_B = 0.55\pm 0.033\mu$K.
 
\section{Discussion}

\subsection{Limits}

As a pilot ALMA study, we reemphasize that the volumes in question are
small, as are the number of detections.  Hence, our conclusions and
uncertainties are dominated by cosmic variance and simple shot noise
(Poisson statistics). Aravena et al. (2016b) consider the issue of
cosmic variance in the context of our particular field. Based on the
drop-out galaxy counts, and the bright submm source counts, this bias
might be as large as a factor two (low). On the other hand,
consideration of the contribution of faint submm continuum sources to
the cosmic infrared background, based on our deeper ASPECS ALMA data,
suggests a factor closer to unity (Aravena et al. 2016b).  Regardless,
since this is a direct survey of the observable in question, namely,
mean brightness due to line emission from distant galaxies at a given
observing frequency, the results remain of interest in general
progress toward millimeter line intensity mapping, and a factor two
uncertainty is inconsequential for our analysis in section 4.2.

Our measurements are also lower limits, since we only sum lines
detected.  We do not extrapolate to, e.g., lower or higher luminosity
galaxies using an assumed luminosity function. Considering CO (the
dominant contributor at 99GHz, certainly, and likely at 242 as well),
our detection threshold was set in order to reach what may be the
'knee' in the CO luminosity function at the primary redshifts to which
our survey is most sensitive ($z \sim 1$ to 3). This estimation was
based on both numerical simulations and extrapolations of CO emission
properties of high redshift galaxies from, e.g., measures of dust
luminosities or star formation rates (see Decarli et al. 2016a for
more details). If the CO luminosity function is relatively flat at low
luminosities, and steep at high luminosities, then galaxies around the
knee of the curve dominate the overall luminosity. For example, using
the Popping et al. (2016) and Lagos et al. (2012) CO luminosity
functions and our limits at 99GHz, we estimate that we should be
detecting between 40\% and 70\% of the total CO luminosity (Decarli et
al. 2016) in this dominant redshift range.

\subsection{Comparison to predictions and CMB spectral distortions}

As stated in Section 1, millimeter line intensity mapping experiments
will have broad impact, from studies of galaxy formation to the Baryon
Acoustic Oscillations. In this section we consider in some detail our
results in the context of one topical area that has seen considerable
attention recently, namely, the spectral distortions of the CMB.

In section 2, we reviewed the predictions for the line brightness at
99GHz and 242 GHZ based on phenomenological calculations using on
proxies for the line luminosity (such as the cosmic star formation
density), or numerical simulations of galaxy formation. Predictions
vary significantly, but range from $\sim 1\mu$K to $10\mu$K, in the
frequency ranges being considered.  To within the
uncertainties inherent in small volume surveys, our direct
measurements of $T_B = 0.94\pm 0.09\mu$K at 99GHz and $T_B = 0.55\pm
0.033\mu$K at 242GHz, argue for the faint end of these predictions,
although we again emphasize that these should be treated as lower
limits.

How do our measurements then compare to, for instance, the expected
distortions in the CMB spectrum due to early energy release, and to
the expected sensitivity of planned CMB spectral distortion
experiments? As a benchmark for experimental sensitivity, we adopt the
current parameters being considered for PIXIE (Kogut et al.  2014;
2011), using the 15 GHz spectral resolution for the proposed
experiment. Considering the expected sky brightness contributions, we
focus on the more cosmologically relevant predictions, relating to
recombination and reionization.

We note that there are other potentially significant foregrounds, in
particular, Galactic and extragalactic thermal emission from warm
dust, and synchrotron emission.  Kogut et al. (2014) review the
relative magnitude of these contributions.  The thermal emission from
warm dust, in particular, is calculated to be an order of magnitude,
or more, stronger than the summed millimeter line emission considered
herein. However, the spectral behavior of the dust emission is
considered to be well understood, and should be well modelled, and
removed, using spectral fitting algorithms over a broad frequency
range.  Herein, we focus on the millimeter and sub-millimeter line
emission, given that this is our measured quantity, and compare it to
the predicted cosmological signals. Additional discussion of
foregrounds can be found in, e.g., de Zotti et al. (2015).

\begin{figure}
\centering 
\includegraphics[width=\columnwidth]{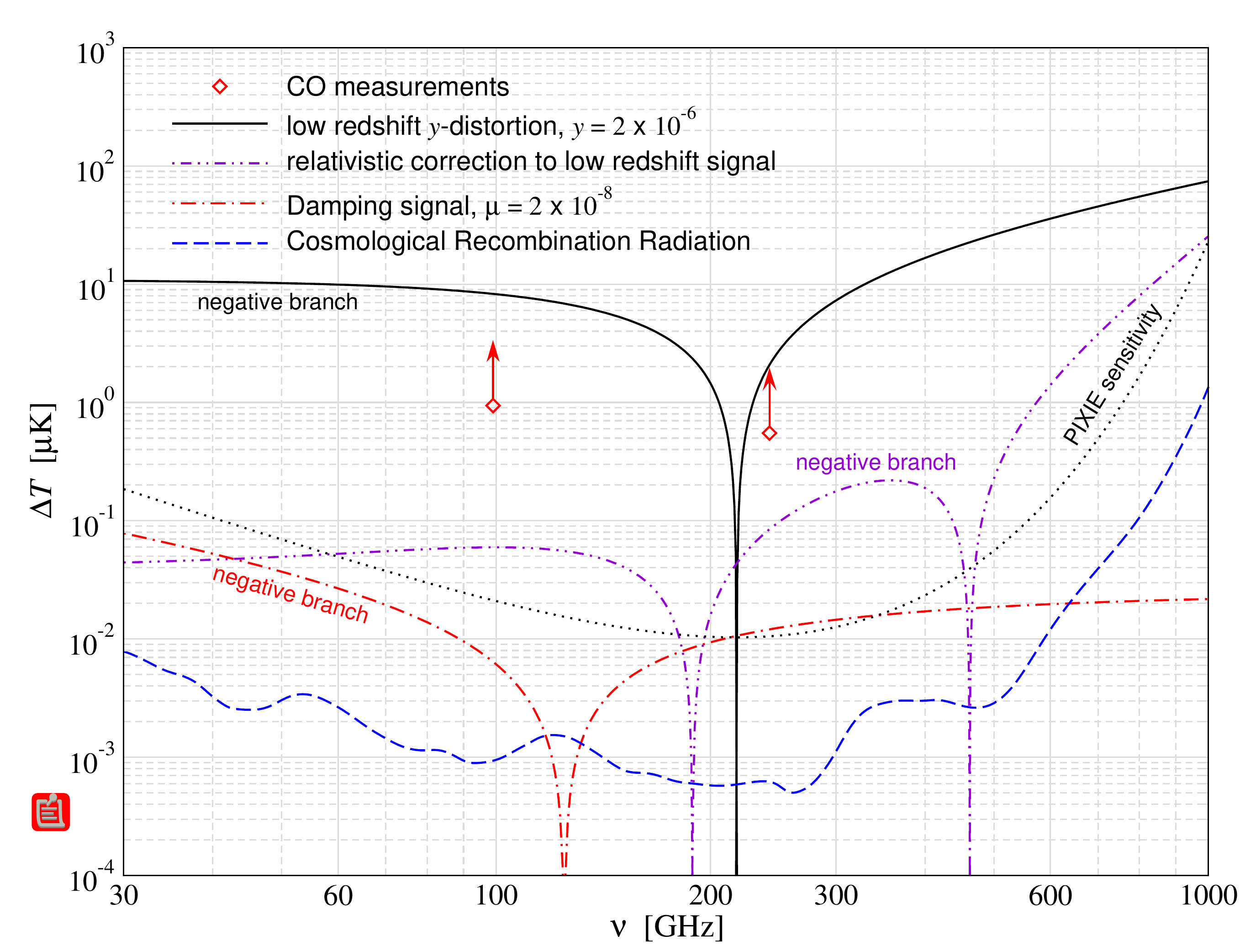}
\caption{Comparison of different CMB distortion signals (negative
branches of the signals are marked) with the millimeter line limits reported
in this paper. The low-redshift distortion created by reionization and
structure formation is close to a $y$-distortion with $y\simeq
2\times 10^{-6}$. Contributions from the hot gas in low mass halos
give rise to a noticeable relativistic temperature correction. For
the damping signal, we plot a $\mu$-distortion with $\mu=2\times
10^{-8}$. The cosmological recombination radiation was computed using
{\tt CosmoSpec}. The estimated effective sensitivity 
($\Delta I_\nu \approx 5\, {\rm Jy~ sr^{-1}}$) 
of PIXIE is shown for
comparison (dotted line).}
\label{fig:one}
\end{figure}

In Fig.~3, we show a comparison of various distortion signals, along
with the line limits derived herein. We focus on {\it guaranteed}
distortions within $\Lambda$CDM (see Chluba, 2016, for most recent
overview), some of which should be detectable with future experiments,
at least in terms of raw sensitivity (Kogut et al. 2011). A wider
range of range of energy release processes (e.g., decaying particle
scenarios) is discussed in Chluba 2013 and Chluba \& Jeong, 2014.

The largest CMB expected spectral distortion is created at
low-redshift by the reionization and structure formation process
(Sunyaev \& Zeldovich, 1972; Hu et al., 1994a). This signal is close
to a pure Compton-$y$ distortion (Zeldovich \& Sunyaev, 1969) caused
through partial up-scattering of CMB blackbody photons by hot
electrons yielding a $y$-parameter $y\simeq 2\times 10^{-6}$ (e.g.,
Refregier et al. 2000; Hill et al., 2015). Contributions from the hot
gas ($\simeq 1{\rm keV}$) residing in low mass halos also give rise to
a noticeable relativistic temperature correction, which could be used
to constrain the average temperature of baryons at low redshifts (Hill
et al., 2015). While the relativistic correction signal requires a
removal of the integrated CO emission, the non-relativistic
$y$-distortion contribution should be less affected and already
separable using multi-frequency capabilities of future experiments.

Another inevitable distortion is created by the dissipation of
small-scale fluctuations in the primordial photon-baryon plasma
(Sunyaev \& Zeldovich, 1970; Daly, 1991; Hu et al., 1994b; Chluba,
Khatri, \& Sunyaev 2012) due to Silk damping.  We illustrate the
$\mu$-distortion (Sunyaev \& Zeldovich, 1970) contribution of this
signal using $\mu=2\times 10^{-8}$, which is close to the value
expected for the $\Lambda$CDM cosmology (Chluba, 2016). A
$\mu$-distortion can only be created in dense and hot environments
present in the early Universe at $z\gtrsim 5\times 10^4$ (Burigana et
al., 1991; Hu \& Silk, 1993). By detecting this signal one can probe
the amplitude of perturbations at scales far smaller than those seen
in the CMB anisotropies, delivering another independent way to test
different inflation models (e.g., Chluba, Khatri, \& Sunyaev 2012;
Chluba, Erickcek \& Ben-Dayan, 2012; Dent et al., 2012; Clesse et al.,
2014).

Finally, we show the cosmological hydrogen and helium recombination
radiation emitted at $z\simeq 10^3$ (Zeldovich et al., 1968; Peebles,
1968; Dubrovich, 1975; Kholupenko et al., 2005;
Rubi{\~n}o-Mart{\'{\i}}n et al., 2006; Chluba \& Sunyaev 2006), which
was computed using {\tt CosmoSpec} (Chluba \& Ali-Ha{\"i}moud,
2016). This signal could provide an independent way to constrain
cosmological parameters and directly map the recombination history
(Sunyaev \& Chluba, 2009). It is unpolarized and its unique spectral
variability is very hard to mimic by other foregrounds or instrumental
effects.\footnote{The expected distortions due to annihilating dark
  matter (McDonald et al., 2001; Chluba 2010; Chluba \& Sunyaev, 2012;
  Chluba 2013) and the differences in the cooling of baryons relative
  to CMB photons during cosmic expansion (Chluba, 2005; Chluba \&
  Sunyaev, 2012) were not illustrated here.}

The latter two effects cause fractional spectral distortions in the
range of 10$^{-9}$ to 10$^{-8}$, implying observed brightness
temperature perturbations $\Delta T_{\rm B}\simeq 3{\rm nK}-30{\rm
nK}$, well below the contribution of the mean line brightness measured
herein. Thus, beyond doubt, an extraction of these primordial
distortions will be very challenging, requiring sophisticated
foreground removal techniques, unprecedented control of systematics,
broad spectral coverage and high sensitivity multi-frequency
capabilities. To successfully remove the integrated millimeter and
submillimeter line emission, it will be advantageous to exploit the
synergies between future CMB distortion measurements and observations
similar to those presented here. Given the importance of the
primordial distortion signals to studies of early-universe physics,
this direction is highly relevant.

As ALMA attains full capability, spectral deep fields will become more
efficient and effective, eventually encompassing areas of tens of
square arcminutes. Our pilot studies have already shown the impact of
such measurements over a broad range of problems in modern
astrophysics and cosmology. In parallel, the Jansky Very Large Array
is exploring similar deep spectral searches at 30GHz (eg. Lentati et
al. 2015; Riechers et al. in prep.), while the advent of high
frequency spectral cameras on the Green Bank Telescope provide a
sensitive platform for wide field spectral searches (Sieth et
al. 2016).  In the long term, a 'Next Generation Very Large Array,'
operating between 20GHz and 115GHz with octave, or broader, bandwidth
receivers and ten times the collecting area of ALMA and the JVLA, has
the potential to revolutionize blind searches for molecular gas in the
early Universe (Carilli et al. 2015; Casey et al. 2015).

\acknowledgements We thank the referee for useful comments that
improved the paper.  FW acknowledges support through ERC grant
COSMIC--DAWN. MA acknowledges partial support from FONDECYT through
grant 1140099. Support for RD was provided by the DFG priority program
1573 `The physics of the interstellar medium'.  DR acknowledges
support from the National Science Foundation under grant number
AST-\#1614213 to Cornell University. FB acknowledges support by the
Collaborative Research Council 956, sub-project A1, funded by the
Deutsche Forschungsgemeinschaft (DFG).  This paper makes use of the
following ALMA data: ADS/JAO.ALMA\#2013.1.00146.S and
ADS/JAO.ALMA\#2013.1.00718.S. ALMA is a partnership of ESO
(representing its member states), NSF (USA) and NINS (Japan), together
with NRC (Canada), NSC and ASIAA (Taiwan), and KASI (Republic of
Korea), in cooperation with the Republic of Chile. The Joint ALMA
Observatory is operated by ESO, AUI/NRAO and NAOJ.  The National Radio
Astronomy Observatory is a facility of the National Science Foundation
operated under cooperative agreement by Associated Universities, Inc.

\clearpage
\newpage

\end{document}